\documentclass[aps,prl,showpacs,twocolumn,groupedaddress]{revtex4}

\usepackage{graphicx}% Include figure files
\usepackage{dcolumn}% Align table columns on decimal point
\usepackage{bm}% bold math

\begin{document}

\preprint{APS/123-QED}

\title{Itinerant-Electron Magnet of the Pyrochlore Lattice: Indium-Doped YMn$_2$Zn$_{20}$}% Force line breaks with \\

%\author{Ann  Author}
% \altaffiliation[Also at ]{Physics Department, XYZ University.}%Lines break aut%omatically or can be forced with \\
%\affiliation{%
%\affiliation{

\author{Yoshihiko Okamoto, Takeshi Shimizu, 
Jun-ichi Yamaura, Yoko Kiuchi, and Zenji Hiroi}
\affiliation{
%$^{1}$
Institute for Solid State Physics, University of Tokyo, Kashiwa 277-8581, Japan
%$^{2}$CREST, Japan Science and Technology Agency (JST), Japan\\
%$^{2}$Department of Advanced Materials, University of Tokyo and CREST-JST, 5-1-5 Kashiwanoha, Kashiwa, Chiba 277-8561, Japan\\
%$^{4}$Institute for Solid State Physics, University of Tokyo, 5-1-5 Kashiwanoha, Kashiwa, Chiba 277-8581, Japan
}

\date{\today}% It is always \today, today,

\begin{abstract}
We report on a ternary intermetallic compound, ``YMn$_2$Zn$_{20}$'', comprising a pyrochlore lattice made of Mn atoms. A series of In-doped single crystals undergo no magnetic long-range order down to 0.4 K, in spite of the fact that the Mn atom carries a local magnetic moment at high temperatures, showing Curie-Weiss magnetism. However, In-rich crystals exhibit spin-glass transitions at approximately 10 K due to a disorder arising from the substitution, while, with decreasing In content, the spin-glass transition temperature is reduced to 1 K. Then, heat capacity divided by temperature approaches a large value of 280 mJ K$^{-2}$ mol$^{-1}$, suggesting a significantly large mass enhancement for conduction electrons. This heavy-fermion-like behavior is not induced by the Kondo effect as in ordinary $f$-electron compounds, but by an alternative mechanism related to the geometrical frustration on the pyrochlore lattice, as in (Y,Sc)Mn$_2$ and LiV$_2$O$_4$, which may allow spin entropy to survive down to low temperatures and to couple with conduction electrons. 
\end{abstract}

\pacs{Valid PACS appear here}% PACS, the Physics and Astronomy
\maketitle

Interesting physics found in some intermetallic compounds containing lanthanide or 
actinide elements results from the interplay between the Kondo effect and the RKKY interaction. 
The former tends to mix $f$-electrons that are localized at high temperatures 
with conduction electrons at low temperatures to form a singlet ground state, 
while the latter stabilizes the magnetic long-range order of $f$-electron spins. 
When the Kondo effect overcomes the RKKY interaction, a heavy-fermion (HF) state 
is generated with the spin entropy of $f$ electrons transferred to conduction electrons, 
resulting in a large mass enhancement. 
A large Sommerfeld constant $\gamma$ of up to several J K$^{-2}$ mol$^{-1}$, 
a large Pauli paramagnetic susceptibility, and $T^2$ resistivity 
with a large coefficient are commonly observed in the HF state.

On the other hand, three transition-metal compounds containing no $f$ electrons 
are known to show similar HF behaviors that may be induced by other mechanisms 
different from the Kondo effect~\cite{1}. 
They are Sc-doped YMn$_2$~\cite{2}, $\beta$-Mn~\cite{3}, and LiV$_2$O$_4$~\cite{4}. 
A cubic-Laves phase YMn$_2$ is an itinerant-electron antiferromagnet with magnetic 
Mn atoms forming a pyrochlore lattice made of corner-sharing tetrahedra. 
It exhibits long-range order with a magnetic moment of 2.7 $\mu_{\textrm{B}}$/Mn 
below 100 K~\cite{5}, but is transformed to a HF state with an enhanced Sommerfeld 
constant of 150 mJ K$^{-2}$ mol$^{-1}$ by substituting 3-5\% Sc for Y~\cite{2}.
$\beta$-Mn has another frustrated lattice consisting of a regular triangle of 
Mn atoms and exhibits a moderate electron mass enhancement~\cite{3,6}. 
LiV$_2$O$_4$ crystallizes in a normal spinel structure, 
where V$^{3.5+}$ ions with the $d^{1.5}$ electron configuration form a pyrochlore lattice. 
It shows a very large $\gamma$ of 420 mJ K$^{-2}$ mol$^{-1}$, 
which is the largest $\gamma$ in $d$ electron compounds~\cite{4,7}. 
Two mechanisms have been proposed to interpret this enhancement for LiV$_2$O$_4$. 
One focuses on a specific aspect of its band structure near the Fermi level~\cite{8}.
According to band-structure calculations, one $d$ electron occupies a narrow $a_{\textrm{1g}}$ 
band and 0.5 electrons form a wide $e_{\textrm{g}}^{\prime}$ band. 
The former tends to be localized on each V atom as if it were an $f$ electron, 
while the latter conduct as $s$ or $p$ electrons, 
as in ordinary HF compounds. 
This situation naturally makes one consider that a type of Kondo scenario is 
applicable to LiV$_2$O$_4$. 
On the other hand, an alternative mechanism sheds light 
on the role of geometrical frustration on the pyrochlore lattice. 
Since spin and/or orbital ordering can be suppressed down to a very 
low temperature by geometrical frustration, 
associated entropy is preserved and dressed up by conduction electrons~\cite{9,10}. 
Although the formation mechanism of the HF state in LiV$_2$O$_4$ still 
remains controversial, the obvious fact that all the $d$-electron HF compounds 
comprise geometrically frustrated lattices made of transition-metal 
atoms strongly suggests that geometrical frustration 
plays a significant role in the formation of the HF state. 

A bottleneck for the investigation of the $d$-electron HF state is the 
limited number of HF compounds. 
Here, we show that a member of ternary intermetallic compounds 
with the general formula $AB_{2}C_{20}$~\cite{10-2, 10-3} can be a good candidate 
for a novel $d$-electron metal with the pyrochlore lattice, 
where $A$ is a rare-earth element, $B$ is a transition-metal element, 
and $C$ is Zn or Al. 
These compounds crystallize in the cubic CeCr$_2$Al$_{20}$ structure 
with the space group $Fd$\={3}$m$~\cite{11}, 
where the $A$ and $B$ atoms occupy the 8a and 16d sites forming 
diamond and pyrochlore lattices, respectively, 
as in the cases of the cubic Laves-phase $AB_2$. 
On the other hand, the $C$ atoms at the 16c, 48f, and 96g sites are located 
between adjacent $A$ atoms, adjacent $B$ atoms, and $A$ and $B$ atoms, 
respectively, as shown in Fig. 1. 
Because of the large population of the $C$ atoms, 
the pyrochlore lattice composed of the $B$ atoms is almost doubly expanded while 
keeping the tetrahedral symmetry in comparison with that in $AB_{2}$. 

In the $AB_{2}C_{20}$ family, compounds with $A$ = Yb, Pr, or U have been extensively studied. 
YbCo$_2$Zn$_{20}$ is particularly focused on, 
because it shows an anomalously large $\gamma$ of about 8 J K$^{-2}$ mol$^{-1}$~\cite{12}. 
Recently, PrIr$_2$Zn$_{20}$ and LaIr$_2$Zn$_{20}$ have been found to be superconductors~\cite{13}. 
In contrast, most compounds with nonmagnetic $A$ atoms such as Y are just Pauli 
paramagnetic metals except for YFe$_2$Zn$_{20}$ that lies in the vicinity 
of ferromagnetism~\cite{14,15}. 
To our knowledge, there are few studies of $AB_2C_{20}$ in terms of the geometrical 
frustration on the pyrochlore lattice of the $B$ atom. 
In this letter, we report on ``YMn$_2$Zn$_{20}$'' comprising a pyrochlore 
lattice composed of the Mn atoms. 
A compound with this ideal composition has not yet been obtained, 
but Benbow and Latturner found in 2006 that the partial 
substitution of In for Zn is effective for stabilizing the compound~\cite{17}. 
They reported the crystal structure of In-doped crystals to be of the CeCr$_2$Al$_{20}$ structure, 
but they did not study their physical properties. 
We prepared a series of single crystals with systematically controlled In content, 
and studied their chemical composition, crystal structure, and physical properties. 
We find a considerably large heat capacity divided by temperature of 
280 mJ K$^{-2}$ mol$^{-1}$ at a low temperature, 
which is almost one order larger than those for 
related compounds and nearly twice as large as that for Sc-doped YMn$_2$. 

A series of single crystals of In-doped YMn$_2$Zn$_{20}$ were prepared by 
the melt-growth method. 
Y, Mn, Zn, and In metals were mixed in a 1 : 2 : 20 $-$ $x_{\textrm{n}}$ : $x_{\textrm{n}}$ molar 
ratio and used to fill an alumina crucible sealed in an evacuated quartz tube. 
The tube was heated to and kept at 1173 K for 24 h, 
cooled to 723 K at a rate of 3 K / h, 
and then furnace-cooled to room temperature. 
The crystals thus prepared are a few mm in size, 
and show metallic luster on \{111\} habit faces. 
The crystal structure was determined by single-crystal and powder X-ray diffraction 
measurements, and the chemical composition was determined by ICP-AES measurements. 
Magnetic susceptibility, specific heat, and resistivity measurements were employed in MPMS and PPMS. 
\begin{table}[tb]
\caption{Chemical compositions and lattice constants of five 
YMn$_{2+\delta}$Zn$_{20-x-\delta}$In$_x$ single crystals. 
The nominal In content $x_{\textrm{n}}$, actual In content $x$, and excess Mn 
content $\delta$ determined by ICP-AES measurements, 
and the lattice parameter $a$ determined at room temperature by powder 
X-ray diffraction measurements are listed. }
\label{t1}

\begin{tabular}{cccc}
& & & \\
\hline
$x_{\textrm{n}}$ & $x$ & $\delta$ & $a$ (\AA) \\
\hline
3 & 2.96(3) & 0.44(2) & 14.4073(8) \\
4 & 3.22(1) & 0.64(2) & 14.417(1) \\
5 & 3.46(1) & 0.65(1) & 14.4659(7) \\
7 & 3.77(2) & 0.99(1) & 14.5024(5) \\
9 & 3.99(2) & 1.24(3) & 14.5273(3) \\
\hline
\end{tabular}
\end{table}

The chemical composition and lattice constant of the five single crystals starting 
from nominal compositions of $x_{\textrm{n}}$ = 3, 4, 5, 7, and 9 are shown in Table I. 
Chemical analyses showed that the actual formula is YMn$_{2+\delta}$Zn$_{20-x-\delta}$In$_x$, 
where the In content $x$ is in the range of 2.96 $\le$ $x$ $\le$ 3.99, and the excess 
Mn content $\delta$ is 0.44 $\le$ $\delta$ $\le$ 1.24. 
Both the decrease in In content from its nominal 
value and the excess Mn content increase with increasing nominal In content. 
The lattice constant $a$ increases linearly as a function of $x$, 
which indicates a systematic variation in chemical composition, 
reflecting the larger metallic radius of In in comparison with Zn. 
The previously reported value of $a$ = 14.7285 \AA \
at $x$ = 5 by Benbow and Latturner 
lies on the linear extrapolation of our data~\cite{17}. 

We have carried out a structural analysis of an $x$ = 3.46 
single crystal, the detail of which will be reported elsewhere. 
According to the results, the 8a, 16d, and 48f sites are fully occupied by Y, Mn, and Zn atoms, 
the same as those in other Y$B_2$Zn$_{20}$ compounds~\cite{10-3}. 
On the other hand, the In atoms are located preferentially at the 16c site, 
and all the remaining Mn, In, and Zn atoms occupy the 96g site randomly. 
The chemical composition at each site is shown in Fig. 1. 
The presence of excess Mn atoms at the 96g site is unfavorable 
for studying the magnetism of the Mn pyrochlore lattice at the 16d site, 
as will be discussed later. Note that $\delta$ increases with 
increasing $x$, so that disturbance by the excess Mn atoms 
is more serious for a larger In content. 

\begin{figure}[tb]
\begin{center}
\includegraphics[width=6.5cm]{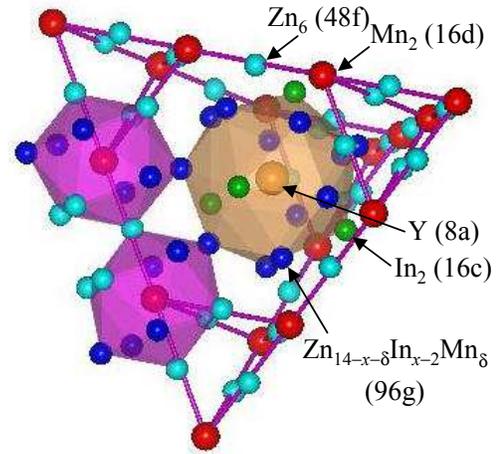}
\end{center}
\caption{(Color online) Crystal structure of YMn$_{2+\delta}$Zn$_{20-x-\delta}$In$_x$. 
Orange, red, green, and sky-blue spheres represent Y (8a), Mn (16d), In (16c), and Zn (48f) atoms, 
respectively. Blue spheres denote the 96g sites that are randomly occupied by Zn, 
In, and Mn atoms with a composition of Zn$_{14-x-\delta}$In$_{x-2}$Mn$_{\delta}$. 
A pyrochlore lattice made of Mn (16d) and coordination polyhedra surrounding 
Y and Mn atoms are also displayed.}
\label{f1}
\end{figure}

The temperature dependence of the magnetic susceptibility $\chi$ of 
YMn$_{2+\delta}$Zn$_{20-x-\delta}$In$_x$ single crystals with 2.96 $\le$ $x$ $\le$ 3.99 
is shown in Fig. 2. 
All the samples clearly show Curie-Weiss behavior at high temperatures, 
indicating that Mn spins are really active as local magnetic moments or 
that a large spin fluctuation exists. 
However, there is no signature for long-range magnetic order in the 
temperature range of 2-300 K. 
A Curie-Weiss fit to the $\chi$ of $x$ = 2.96 with the smallest $\delta$ gives 
an effective moment of $\mu_{\textrm{eff}}$ = 2.4 $\mu_{\textrm{B}}$/Mn and a 
Weiss temperature of $\theta_{\textrm{W}}$ = $-$28 K. 
This $\mu_{\textrm{eff}}$ is comparable to that expected 
when each Mn atom carries an $S$ = 1 localized spin. 
The negative $\theta_{\textrm{W}}$ suggests predominant antiferromagnetic interactions 
that are much weaker than those in YMn$_2$~\cite{5}. 
This must be because of the larger Mn-Mn distance, e.g., 5.11 \AA \ for $x$ = 3.46, 
than 2.72 \AA \ for YMn$_2$.

\begin{figure}[tb]
\begin{center}
\includegraphics[width=8cm]{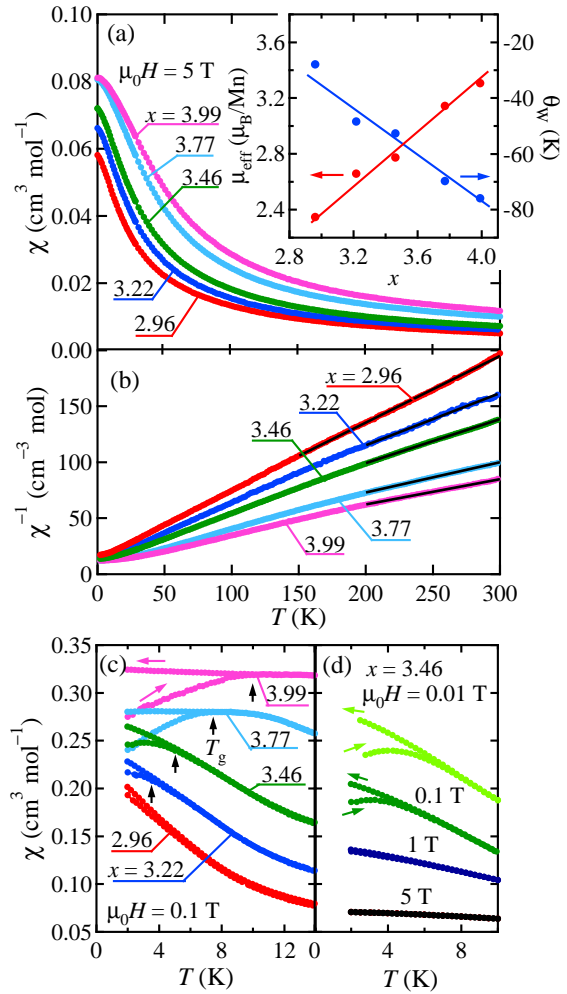}
\end{center}
\caption{(Color online) Temperature dependences of magnetic susceptibility $\chi$ (a) 
and its inverse (b) for YMn$_{2+\delta}$Zn$_{20-x-\delta}$In$_x$ (2.96 $\le$ $x$ $\le$ 3.99) 
single crystals measured on heating in a magnetic field of 5 T. 
The solid line on each $\chi^{-1}$ data in (b) represents a Curie-Weiss fit. 
The inset of (a) shows the $x$ dependences of the effective moment $\mu_{\textrm{eff}}$ and 
the Weiss temperature $\theta_{\textrm{W}}$ obtained from the Curie-Weiss fit. 
(c) Temperature dependences of field-cooled and zero-field-cooled $\chi$ measured 
at a magnetic field of 0.1 T. 
For clarity, the curves are shifted by 0.03, 0.06, 0.13, and 0.18 cm$^3$ mol$^{-1}$ for 
$x$ = 3.22, 3.46, 3.77 and 3.99, respectively. 
The arrow denotes a spin glass transition temperature $T_{\textrm{g}}$. 
(d) Temperature dependences of field-cooled and zero-field-cooled $\chi$ measured at 
various magnetic fields of up to 5 T for $x$ = 3.46.}
\label{f2}
\end{figure}

As $x$ increases, the $\chi$ data shift upward, 
and $\mu_{\textrm{eff}}$ and $|\theta_{\textrm{W}}|$ increase, 
as shown in the inset of Fig. 2 (a). 
This is possibly due to a growing contribution of the excess Mn atom at the 96g site, 
which may have a larger effective moment interacting with neighboring moments 
at the 16d site more strongly: the distance between the 96g and 16d sites is 2.8 \AA, 
nearly half of the 16d-16d distance (Fig. 1). 

The $\chi$ data for a large $x$ show a distinct thermal hysteresis between 
zero-field-cooling (ZFC) and field-cooling processes at a low magnetic field of 0.1 T, 
as shown in Fig. 2 (c). 
The hysteresis disappears at larger magnetic fields such as 1 T, 
as shown in Fig. 2 (d) for $x$ = 3.46. 
Thus, a spin glass transition must take place. 
The transition temperature $T_{\textrm{g}}$, 
defined as the temperature below which a hysteresis starts to grow, is 10 K for $x$ = 3.99; 
it gradually decreases with decreasing $x$ and becomes below 2 K for $x$ = 2.96. 
It is plausible to ascribe the origin of this spin glass transition
to the excess Mn atoms existing at the 96g site, 
because they must bring a certain disorder into the magnetic network on the pyrochlore lattice. 
The observed reduction in $T_{\textrm{g}}$ with decreasing $x$ may be 
due to the decrease in $\delta$. 
A similar disorder-induced spin glass has been found in Al-doped YMn$_2$~\cite{18}. 

\begin{figure}[tb]
\begin{center}
\includegraphics[width=8cm]{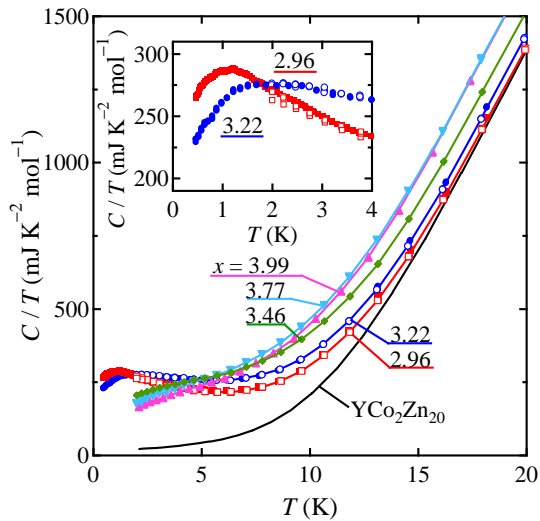}
\end{center}
\caption{(Color online) Heat capacity divided by temperature $C$ / $T$ vs 
$T$ plot for YMn$_{2+\delta}$Zn$_{20-x-\delta}$In$_x$ single crystals. 
The data for a polycrystalline sample of YCo$_2$Zn$_{20}$ are also shown for comparison. 
The inset shows an expanded view of the low-temperature part of $C$ / $T$ for $x$ = 2.96 and 3.22.}
\label{f3}
\end{figure}

No sharp peak indicative of a magnetic or structural transition has been observed 
in the temperature dependence of heat capacity for any samples in the range between 
0.4 or 2 and 300 K. 
Figure 3 shows the heat capacity divided by temperature as a function of $T$ below 20 K. 
$C$ / $T$ is large, compared with that of the Pauli paramagnetic YCo$_2$Zn$_{20}$, 
revealing that large spin entropy remains at a low temperature for the Mn compound. 
For $x$ = 2.96, $C$ / $T$ starts to deviate from the data of YCo$_2$Zn$_{20}$ below 15 K 
and increases gradually with further cooling, followed by a broad maximum 
at $T_{\textrm{max}}$ = 1.2 K. 
The maximum $C$ / $T$ is 280 mJ K$^{-2}$ mol$^{-1}$. 
As $x$ increases, the peak becomes smaller and shifts to higher temperatures: 
$T_{\textrm{max}}$ = 2 and 4 K for $x$ = 3.22 and 3.46, respectively. 
This trend is similar to what is observed in the $\chi$ data of Fig. 2 (c), 
where a broad peak in the ZFC data at low magnetic fields moves 
to higher temperatures with increasing $x$. 
In fact, both the two peaks in $C$ / $T$ and ZFC $\chi$ appear at 4 K for $x$ = 3.46. 
Therefore, the observed broad peak in $C$ / $T$ must be related to the 
spin glass transition. 
Probably, a part of the spin degree of freedom has been frozen 
through the spin glass transition, giving rise to a downturn 
in $C$ / $T$ below $T_{\textrm{g}}$. 

A significant contribution from the excess Mn spins must be 
included in the present $C$ / $T$ data, particularly for 
large $x$ ($\delta$) values and at high temperatures. 
Nevertheless, the fact that $C$ / $T$ at the lowest temperature 
increases with decreasing $x$, that is, with decreasing contribution 
of the excess Mn spins, implies that there is a large, 
intrinsic $T$-linear term in the heat capacity of ideal YMn$_2$Zn$_{20}$. 
One expects that $C$ / $T$ at $T$ = 0 would approach even a larger 
value without a saturation associated with the spin glass transition 
in ``YMn$_2$Zn$_{20}$''. 
Provided that this $T$-linear term comes from conduction electrons, 
as in conventional metallic compounds, the Sommerfeld constant $\gamma$ of
``YMn$_2$Zn$_{20}$'' can be large, i.e., more than 280 mJ K$^{-2}$ mol$^{-1}$, 
which is much larger than $\gamma$ = 18 mJ K$^{-2}$ mol$^{-1}$ for 
YCo$_2$Zn$_{20}$ and 53 mJ K$^{-2}$ mol$^{-1}$ for nearly ferromagnetic 
YFe$_2$Zn$_{20}$~\cite{15}. 
Since band-structure calculations for YCo$_2$Zn$_{20}$ and YFe$_2$Zn$_{20}$ give 
$\gamma_{\textrm{band}}$ = 19 and 37 mJ K$^{-2}$ mol$^{-1}$, respectively, 
the mass enhancement factor $f$ = $\gamma$ / $\gamma_{\textrm{band}}$ is equal 
to 1.0 for YCo$_2$Zn$_{20}$ and to 1.4 for YFe$_2$Zn$_{20}$~\cite{15}. 
For ``YMn$_2$Zn$_{20}$'', a preliminary band-structure calculation 
by Harima gives $\gamma_{\textrm{band}}$ = 34 mJ K$^{-2}$ mol$^{-1}$~\cite{19}, 
so that one obtains a large mass enhancement factor of $f$ = 8.2, 
using the maximum $C$ / $T$ for $x$ = 2.96 as $\gamma$. 
It is likely that electron correlation effects are enhanced in the Mn compound, 
because Mn eventually favors a half-filled $d$ band in a crystal.

\begin{figure}[tb]
\begin{center}
\includegraphics[width=7cm]{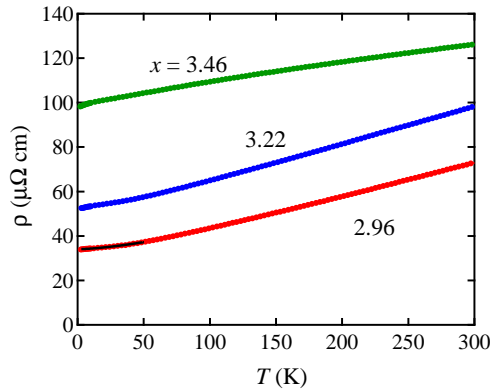}
\end{center}
\caption{(Color online) Temperature dependence of electrical 
resistivity $\rho$ of YMn$_{2+\delta}$Zn$_{20-x-\delta}$In$_x$ single crystals 
measured with current running along the [110] direction. 
A solid curve on the $x$ = 2.96 data in the range 3-100 K represents a fit 
to the form $\rho$ = $\rho_0$ $+$ $AT^{n}$, where $\rho_0$ = 34.0 $\mu$$\Omega$ cm, $n$ = 1.54 
and $A$ = 8.0 $\times$ 10$^{-3}$ $\mu$$\Omega$ cm K$^{-1.54}$.}
\label{f4}
\end{figure}

When the effective mass of conduction electrons is increased, 
the coefficient of the $T^2$ term in resistivity should become large. 
The resistivity of $x$ = 2.96 is shown in Fig. 4, 
which exhibits an almost linear temperature dependence at high temperatures 
and approaches a large residual value $\rho_0$ as $\rho$ = $\rho_0$ $+$ $AT^{n}$: 
$\rho_0$ = 34.0 $\mu$$\Omega$ cm, $n$ = 1.54 
and $A$ = 8.0 $\times$ 10$^{-3}$ $\mu$$\Omega$ cm K$^{-1.54}$. 
The power is considerably smaller than 2 expected for a Fermi liquid, 
but is close to the value of 3/2 predicted to appear in the vicinity 
of an antiferromagnetic quantum critical point 
in a three-dimensional system by the spin-fluctuation theory~\cite{20}. 
We have to consider, however, additional contributions of magnetic scattering 
by extra Mn atoms at the 96g site and also those of lattice imperfections caused by In doping. 
In fact, the residual resistivity is considerably large, 
resulting in a small residual resistivity ratio (RRR) of $\sim$ 2. 
Larger $\rho_0$ and smaller RRR values for $x$ = 3.22 and 3.46 may reflect 
larger contributions of such unfavorable effects. 
A similar behavior in resistivity has been observed for (Y,Sc)Mn$_2$, 
when disorder is introduced by Al doping to the Mn site~\cite{21}. 
Future experiments using better samples with less disorder 
would clarify the inherent resistivity of ``YMn$_2$Zn$_{20}$''.

Geometrically frustrated systems often suffer from disorder, 
which tends to mask intrinsic properties. 
This is because even a small amount of disorder can seriously influence 
the surroundings at $T$ = 0 in the absence of a long-range order. 
In our cleanest sample of $x$ = 2.96, the spin degree of freedom is partly frozen 
owing to a magnetic disorder induced by the excess Mn atom. 
If pure ``YMn$_2$Zn$_{20}$'' comprising a perfect Mn pyrochlore lattice is obtained, 
one expects a larger mass enhancement or an unknown exotic ground state 
associated with the geometrical frustration. 
There must be interesting physics in the itinerant-electron antiferromagnet 
on the pyrochlore lattice. 

In conclusion, we find that In-doped YMn$_2$Zn$_{20}$ is a novel strongly 
correlated electron system with a pyrochlore lattice, 
where the $T$-linear term in heat capacity is enormously enhanced 
to 280 mJ K$^{-2}$ mol$^{-1}$. 
This enhancement is obviously not attributed to the Kondo effect, 
but to the geometrical frustration on the pyrochlore lattice. 

%\section*{Acknowledgment}

We thank H. Harima for drawing our attention to the present compound 
and also for informing us of his preliminary calculations on the band structure. 
This work was partly supported by a Grant-in-Aid for 
Scientific Research on Priority Areas ``Novel States of Matter Induced by Frustration''
 (No. 19052003) provided by MEXT, Japan.

%\bibliography{aps9.bib}% Produces the bibliography via BibTeX.

\begin{thebibliography}{99}

\bibitem{1} C. Lacroix: J. Phys. Soc. Jpn. \textbf{79} (2010) 011008.
\bibitem{2} H. Wada
	%, H. Nakamura, E. Fukami, K. Yoshimura, M. Shiga, and Y. Nakamura
	\textit{et al.}%
	: J. Magn. Magn. Mat. \textbf{70} (1987) 17.
\bibitem{3} T. Shinkoda, K. Kumagai, and K. Asanuma: 
	J. Phys. Soc. Jpn. \textbf{46} (1979) 1754.
\bibitem{4} S. Kondo
	\textit{et al.}%
	%, D. C. Johnston, C. A. Swenson, F. Borsa, A. V. Mahajan, L. L. Miller, T. Gu, 
	%A. I. Goldman, M. B. Maple, D. A. Gajewski, E. J. Freeman, N. R. Dilley, R. P. Dickey, 
	%J. Merrin, K. Kojima, G. M. Luke, Y. J. Uemura, O. Chmaissem, and J. D. Jorgensen: 
	: Phys. Rev. Lett. \textbf{78} (1997) 3729.
\bibitem{5} Y. Nakamura, M. Shiga, and S. Kawano: 
	Phys. B \textbf{120} (1983) 212.
\bibitem{6} H. Nakamura
	\textit{et al.}%
%	, K. Yoshimoto, M. Shiga, M. Nishi, and K. Kakurai
	: J. Phys.: Condens. Matter \textbf{9} (1997) 4701.
\bibitem{7} C. Urano
	\textit{et al.}%
%	, M. Nohara, S. Kondo, F. Sakai, H. Takagi, T. Shiraki, and T. Okubo: 
	: Phys. Rev. Lett. \textbf{85} (2000) 1052.
\bibitem{8} V. I. Anisimov
	\textit{et al.}%
%	, M. A. Korotin, M. Z\"{o}lfl, T. Pruschke, K. Le Hur, and T. M. Rice
	: Phys. Rev. Lett. \textbf{83} (1999) 364.
\bibitem{9} V. Eyert
	\textit{et al.}%
%	, K.-H. H\"{o}ck, S. Horn, A. Loidl, and P. S. Riseborough
	: Europhys. Lett. \textbf{46} (1999) 762.
\bibitem{10} Y. Yamashita and K. Ueda: 
	Phys. Rev. B \textbf{67} (2003) 195107.
\bibitem{10-2} S. Niemann and W. Jeitschko: 
	J. Solid State Chem. \textbf{114} (1995) 337.
\bibitem{10-3} T. Nasch, W. Jeitschko, and U. C. Rodewald: 
	Z. Naturforsch B \textbf{52} (1997) 1023.
\bibitem{11} P. I. Kripyakevich and O. S. Zarechnyuk: Dopov. Akad. Nauk Ukr. RSR, Ser. A 
	\textbf{30} (1968) 364.
\bibitem{12} M. S. Torikachvili
	%S. Jia, E. D. Mun, S. T. Hannahs, R. C. Black, W. K. Neils, D. Martien, S. L. Bud'ko, 
	%and P. C. Canfield%
	\textit{et al.}%
	: Proc. Natl. Acad. Sci. U.S.A. \textbf{104} (2007) 9960.
\bibitem{13} T. Onimaru
	%, K. T. Matsumoto, Y. F. Inoue, K. Umeo, Y. Saiga, Y. Matsushita, 
	%R. Tamura, K. Nishimoto, I. Ishii, T. Suzuki, and T. Takabatake%
	\textit{et al.}%
	: J. Phys. Soc. Jpn. \textbf{79} (2010) 033704.
\bibitem{14} S. Jia
	\textit{et al.}%
%	, S. L. Bud'ko, G. D. Samolyuk, and P. C. Canfield%
	: Nat. Phys. \textbf{3} (2007) 334.
\bibitem{15} S. Jia
	\textit{et al.}%
%	, N. Ni, G. D. Samolyuk, A. Safa-Sefat, K. Dennis, H. Ko, G. J. Miller, 
%	S. L. Bud'ko, and P. C. Canfield%
	: Phys. Rev. B \textbf{77} (2008) 104408.
\bibitem{17} E. M. Benbow and S. E. Latturner: J. Solid State Chem. \textbf{179} (2006) 3989.
\bibitem{18} K. Motoya: J. Phys. Soc. Jpn. \textbf{55} (1986) 3733.
\bibitem{19} H. Harima: private communication.
\bibitem{20} T. Moriya: \textit{Spin Fluctuation in Itinerant Electron Magnetism} (Springer,
	 Berlin, 1985).
\bibitem{21} M. Shiga, K. Fujisawa, and H. Wada: 
	J. Phys. Soc. Jpn. \textbf{62} (1993) 1329.

\end{thebibliography}

\end{document}